\documentclass[proof]{WileyASNA-v1}

\articletype{Article Type}

\received{}
\revised{}
\accepted{}

\newcommand{\ud}{\mathrm{d}}
\newcommand{\pd}{\partial}

\newcommand{\lie}{\mathscr L}

\newcommand{\order}[1]{\mathcal O\left(#1\right)}
\newcommand{\scri}{\mathscr I}

\begin{document}

\title{Gravitational memory effects  in Brans-Dicke theory}

\author[1]{Shaoqi Hou*}

\authormark{SHAOQI HOU}

\address[1]{\orgdiv{School of Physics and Technology}, \orgname{Wuhan University}, \orgaddress{\state{Hubei}, \country{China}}}

\corres{*  \email{hou.shaoqi@whu.edu.cn}}

\presentaddress{School of Physics and Technology, Wuhan University, Wuhan, Hubei 430072, China}

\abstract{There exist gravitational memory effects in Brans-Dicke theory.
They are closely related to the Bondi-Metzner-Sachs symmetries living on the null infinity in an isolated system.
By studying the asymptotically flat spacetime in Brans-Dicke theory and the asymptotic symmetries, one discovers that the displacement memory effect in the tensor sector is due to the vacuum transition caused by the null energy fluxes penetrating the null infinity, while in the scalar sector, the vacuum transition is due to the angular momentum fluxes. 
Together with the spin and the center-of-mass memory effects, the displacement memories are constrained by various flux-balance laws.}



\maketitle

\footnotetext{\textbf{Abbreviations:} GR, general relativity; BD, Brans-Dicke theory; GW, gravitational wave; BMS, Bondi-Metzner-Sachs} 

\section{Introduction}
\label{sec-intro}

Gravitational memory effects are fascinating phenomena that occur in general relativity (GR) (\cite{Zeldovich:1974gvh,Braginsky:1986ia,Christodoulou1991,Thorne:1992sdb}).
There are several types of memories.
The displacement memory refers to the permanent change in the relative distance between two test particles after the passage of gravitational waves (GWs).
The spin memory manifests itself in the different propagating times it takes for two test particles, orbiting in a circle in opposite directions, to return to their original positions, when GWs pass by (\cite{Pasterski:2015tva}).
And the center-of-mass (CM) memory is the lasting shift of the CM of an isolated system due to GWs (\cite{Nichols:2018qac}).
There are other memories, e.g., velocity memory effect (\cite{Zhang:2017rno,Zhang:2017geq,Zhang:2018srn}), none of which will be discussed in this work.
Both displacement and spin memories might be detected by interferometers and pulsar timing arrays (\cite{Seto:2009nv,Wang2015mm,Lasky:2016knh,McNeill:2017uvq,Johnson:2018xly,Hubner2020mmn,Boersma:2020gxx,Madison:2020xhh}), but it is difficult to observe the CM memory experimentally (\cite{Nichols:2018qac}).

Memory effects might also exist in modified theories of gravity (\cite{Lang:2013fna,Lang:2014osa,Du:2016hww,Kilicarslan:2018bia,Kilicarslan:2018yxd,Kilicarslan:2018unm}).
Brans-Dicke (BD) theory is the simplest modified gravity (\cite{Brans:1961sx}), whose action is 
\begin{equation}
    \label{eq-act-bd}
    S=\frac{1}{16\pi G_0}\int\ud^4 x\sqrt{-g}\left( \varphi R-\frac{\omega}{\varphi}\nabla_a\varphi\nabla^a\varphi \right),
\end{equation}
where $\omega$ is a constant, $G_0$ is the bare gravitational constant.
The memory effect in this theory would be the topic of this work.
It is found out that there are not only the same memories in BD as those in GR, but also a new memory effect due to the BD scalar field, named S memory in \cite{Du:2016hww}. 
These memories also have something to do with the asymptotic symmetries of an isolated system, the so-called Bondi-Metzner-Sachs (BMS) group,
which is a semi-direct product of the  supertranslation group and the Lorentz group as in GR. 
On the one hand, the supertranslation causes the vacuum transition in the tensor sector, and is responsible for the displacement memory, while the Lorentz transformation switches vacua in the scalar sector, so the S memory takes place.
On the other hand, the BMS symmetries imply the existence of supermomenta, angular momenta and Lorentz boost charges at the null infinity $\scri$.
When there are GWs, these quantities vary over time, and their variations equal the corresponding fluxes escaping from the source of gravity.
The flux-balance laws provide constraints on the memory effects.
For more details, please refer to \cite{Hou:2020tnd,Tahura:2020vsa,Hou:2020wbo}.

This work mainly discusses the displacement memories in the tensor and the scalar sectors.
Their relations with the BMS symmetries will be deciphered.
Finally, we will compute the constraints on various memory effects using the flux-balance laws.
We hope this work will set up the basis for using memory effects to probe the nature of gravity (\cite{Koyama:2020vfc}).
Recently, the properties of gravity has been studied by the scattering of Dirac spinors and the scalar field \cite{Gao:2019tde,Fu:2020vyo}.
So this work is organized in the following way.
Section~\ref{sec-iso} discusses the asymptotically flat spacetime in BD. 
Based on that, the BMS symmetry is defined and computed in Sec.~\ref{sec-bms}, and then, the ``conserved charges'' and fluxes are determined in Sec.~\ref{sec-ccf}.
Section~\ref{sec-mem} focuses on memory effects.
There, the displacement memories are analyzed in Sec.~\ref{sec-dis}, and Sec.~\ref{sec-spin} is devoted to the spin and the CM memories.
In the end, there is a short summary in Sec.~\ref{sec-con}.
In this work, the abstract index notation is used (\cite{Wald:1984rg}), and the speed of light in vacuum is $c=1$.

\section{Asymptotically flat spacetimes}
\label{sec-iso}

Memory effects also happen in cosmological background (\cite{Donnay:2019zif,Bonga:2020fhx}), but here, we will focus on those occuring in an isolated system. 
At the distances very far away from the source of gravity, the spacetime is nearly Minkowskian, so such kind of spacetime is said to be asymptotically flat.
An  asymptotically flat spacetime at null infinity in GR can be defined in a coordinate independent manner; see \cite{Wald:1984rg}. 
In BD, one can also propose a similar definition as presented in \cite{Hou:2020wbo}. 
But here, we will simply define such a spacetime using the generalized Bondi-Sachs coordinates $\{u,r,\theta,\phi\}$ ($x^A=\theta,\phi$), in which the metric is 
\begin{equation}
    \label{eq-bc}
    \ud s^2=g_{uu}\ud u^2+2g_{ur}\ud u\ud r+2g_{uA}\ud u\ud x^A+h_{AB}\ud x^A\ud x^B.
\end{equation}
One requires that the metric components satisfy the following boundary conditions,
\begin{subequations}
    \label{eq-exp-bd}
    \begin{gather}
        g_{uu}=-1+\order{r^{-1}},\quad g_{ur}=-1+\order{r^{-1}},\label{eq-exp-bd-1}\\
        g_{uA}=\order{1},\quad h_{AB}=r^2\gamma_{AB}+\order{r},\label{eq-exp-bd-2}
    \end{gather}
\end{subequations}
where  $\gamma_{AB}$ is the round metric on a unit 2-sphere, 
\begin{equation}
    \label{eq-gmet}
    \gamma_{AB}\ud x^A\ud x^B=\ud\theta^2+\sin^2\theta\ud\phi^2.
\end{equation}
The scalar field behaves like $\varphi=\varphi_0+\order{r^{-1}}$ with $\varphi_0$ a constant.
One also imposes the determinant condition, 
\begin{equation}
    \label{eq-det}
    \det(h_{AB})=r^4\left(  \frac{\varphi_0}{\varphi}\right)^2\sin^2\theta,
\end{equation}
so that $r$ becomes the luminosity distance as it approaches the null infinity.

With these conditions, one can obtain the series solutions to the equations of motion, i.e.,
\begin{subequations}
    \label{eq-sol}
    \begin{eqnarray}
    \varphi&=&\varphi_0+\frac{\varphi_1}{r}+\frac{\varphi_2}{r^2}+\order{\frac{1}{r^3}},\label{eq-exp-s}\\
        g_{uu}&=&-1+\frac{2m+\varphi_1/\varphi_0}{r}+\order{\frac{1}{r^2}},\\
     g_{ur}&=&-1+\frac{\varphi_1}{\varphi_0r}+\frac{1}{r^2}\left[ \frac{1}{16}\hat c_{A}^{B}\hat c^{A}_{B}+\frac{2\omega-5}{8}\left( \frac{\varphi_1}{\varphi_0} \right)^2+\frac{\varphi_2}{\varphi_0} \right]\nonumber\\
     &&+\order{\frac{1}{r^3}},\\ 
         g_{uA}&=&\frac{\mathscr D_B\hat c^B_A}{2}+\frac{2}{3r}\left[ N_A+\frac{1}{4}\hat c_{AB}\mathscr D_C\hat c^{BC}-\frac{\varphi_1}{12\varphi_0}\mathscr D_B\hat c^B_A \right]\nonumber\\
         &&+\order{\frac{1}{r^2}},\\
     g_{AB}&=&r^2\gamma_{AB}+r\left( \hat c_{AB}-\gamma_{AB}\frac{\varphi_1}{\varphi_0} \right)+\hat d_{AB}\nonumber\\
     &&+\gamma_{AB}\left( \frac{1}{4}\hat c_{C}^{D}\hat c^{C}_{D}+\frac{\varphi_1^2}{\varphi_0^2}-\frac{\varphi_2}{\varphi_0} \right)+\order{\frac{1}{r}}.
    \end{eqnarray}
\end{subequations}
Here, $\varphi_1,\,\varphi_2,\,\hat c_{AB}$, and $\hat d_{AB}$ are functions of $u$ and $x^A$ with $\gamma^{AB}\hat c_{AB}=\gamma^{AB}\hat d_{AB}=0$, and $\mathscr D_A$ is the covariant derivative for $\gamma_{AB}$. 
$m=m(u,x^A)$ and $N_A=N_A(u,x^B)$ are called the Bondi mass and angular momentum aspects, respectively.
Their evolution equations are
\begin{subequations}
\begin{eqnarray}
    \label{eq-uu2-mdot}
    \dot {m}&=&-\frac{\mathscr D_A\mathscr D_BN^{AB}}{4}-\frac{N_{AB}N^{AB}}{8}-\frac{2\omega+3}{4}\left( \frac{N}{\varphi_0} \right)^2,\\
    \dot {N}_A&=&\mathscr D_Am
        +\frac{1}{4}(\mathscr D_B\mathscr D_A\mathscr D_C\hat c^{BC}-\mathscr D_B\mathscr D^B\mathscr D_C\hat c_A^C)\nonumber\\ 
        &&-\frac{1}{16}\mathscr D_A(N^{B}_{C}\hat c_{B}^{C})+\frac{1}{4}N^{B}_{C}\mathscr D_A\hat c_{B}^{C}\nonumber\\
        &&+\frac{1}{4}\mathscr D_B(N_{A}^{C}\hat c^{B}_{C}-\hat c_{A}^{C}N^{B}_{C})\nonumber\\ 
        &&+\frac{2\omega+3}{8\varphi_0^2}(\varphi_1\mathscr D_AN-3N\mathscr D_A\varphi_1),\label{eq-uA2-ndot}
\end{eqnarray}
\end{subequations}
where $N_{AB}=-\pd\hat c_{AB}/\pd u$ is the news tensor, and $N=\pd\varphi_1/\pd u$.
When there are no GWs, both $N_{AB}$ and $N$ vanish. 
So they are called the radiative degrees of freedom.
In particular, $N_{AB}$ corresponds to the tensor GW, while $N$ to the scalar GW (\cite{Hou:2017bqj}).

At the null infinity, although the spacetime is very similar to the flat one, the symmetry group living there is not simply the Poincar\'e group.
It is a much larger group, instead.
In the next subsection, this group will be analyzed.
After that, the ``conserved charges'' and fluxes will be obtained.

\subsection{Bondi-Metzner-Sachs symmetries}
\label{sec-bms}

The asymptotic symmetries, or BMS symmetries, are diffeomorphisms that preserve the boundary conditions defined in the previous section.
An infinitesimal asymptotic symmetry $\xi^a$ is the easiest to be determined by solving the following conditions (\cite{Hou:2020tnd})
\begin{subequations}
    \begin{gather}
        \lie_\xi g_{rr}=\lie_\xi g_{rA}=0,\label{eq-bms-inf-0}\\
         g^{AB}\lie_\xi g_{AB}=-\frac{2\lie_\xi\varphi}{\varphi},\label{eq-bms-inf-1}\\
        \lie_\xi g_{ur}=\order{r^{-1}},\; \lie_\xi g_{uA}=\order{1},\; \lie_\xi g_{AB}=\order{r},\label{eq-bms-inf-2}\\ 
        \lie_\xi g_{uu}=\order{r^{-1}},\label{eq-bms-inf-3}\\
        \lie_\xi\varphi=\order{r^{-1}}.\label{eq-bms-inf-s}
    \end{gather}
\end{subequations}
The result shows that $\xi^a$ is parameterized by two functions $\alpha(x^A)$ and $Y^A(x^B)$, and explicitly given by
\begin{subequations}
    \begin{gather}
        \xi^u=f,\label{eq-xi-u}\\ 
        \xi^A=Y^A-\frac{\mathscr D^Af}{r}+\order{\frac{1}{r^2}},\\ 
        \xi^r=-\frac{r}{2}\psi+\frac{1}{2}\mathscr D^2f+\order{\frac{1}{r}}.
    \end{gather}
\end{subequations}
with $f=\alpha+u\psi/2$ and $\psi=\mathscr D_AY^A$.
The symmetries parameterized by $\alpha$ are called supertranslations.
If $\alpha=\alpha_0+\alpha_1\sin\theta\cos\phi+\alpha_2\sin\theta\sin\phi+\alpha_3\cos\theta$ with $\alpha_i\, (i=0,1,2,3)$ constant, one obtains the usual translation. 
A generic supertranslation is described by an arbitrary linear combination of spherical harmonics $Y_{lm}$, so the group of supertranslations, the supertranslation group, is infinitely dimensional.
The (global) conformal Killing vector field $Y^A$, satisfying $\lie_Y\gamma_{AB}=\psi\gamma_{AB}$, determines an infinitesimal Lorentz transformation.
The BMS group is thus the semi-direct product of the supertranslation group and the Lorentz group as in GR.
In particular, the supertranslation group is abelian and the normal subgroup of the BMS group.

The BMS group is much larger than the Poincar\'e group, as the former is infinitely dimensional, while the later has the dimension of ten.
In fact, the BMS group contains an infinite copies of the Poincar\'e group, which are related to each other through the supertranslation.

\subsection{``Conserved charges'' and fluxes}
\label{sec-ccf}

By Noether's theorem, one may define conserved quantities for the BMS symmetries. 
However, when there are GWs, none of these quantities is exactly a constant of time ($u$).
Their changes are given by the corresponding fluxes. 

With the Hamiltonian formalism designed by \cite{Wald:1999wa}, the charges and fluxes can be calculated (\cite{Hou:2020wbo}).
For a supertranslation, one finds the supermomentum,
                \begin{equation}
                    \label{eq-smc}
                    \mathcal P_\alpha[\mathscr C]=\frac{\varphi_0}{4\pi G_0}\oint_{\mathscr C}\alpha m\ud^2\boldsymbol{\Omega},
                \end{equation}
where $\mathscr C$ represents a cross section on $\scri$, and $\ud^2\boldsymbol{\Omega}=\sin\theta\ud\theta\ud\phi$.
If $\alpha=1$, one has the Bondi mass, and if $\alpha$ is a linear combination of $Y_{lm}$ with $l=1$, one has the spatial momentum.
The flux is
            \begin{eqnarray}
                \label{eq-falpha}
                F_\alpha[\mathscr B]=\frac{\varphi_0}{16\pi G_0}\int_{\mathscr B}&&\alpha\Bigg[  \mathscr D_A\mathscr D_BN^{AB}+\frac{1}{2}N_A^BN^A_B\nonumber\\
                &&+(2\omega+3)\left( \frac{\varphi_1}{\varphi_0} \right)^2 \Bigg]\ud u\ud^2\boldsymbol{\Omega},
            \end{eqnarray}
where $\mathscr B$ is a subset of $\scri$. 
The first term in the square brackets represents the so-called soft part, and the remaining are the hard part.
If $\mathscr C$ and $\mathscr C'$ are the past and the future boundaries of $\mathscr B$, the flux-balance law is 
\begin{equation}
    \label{eq-fb-a}
F_\alpha[\mathscr B]=-(\mathcal P_\alpha[\mathscr C']-\mathcal P_\alpha[\mathscr C]).
\end{equation}
This will be useful for constraining the displacement memory in the tensor sector.

For an infinitesimal Lorentz transformation,  one writes $Y^A=\mathscr D^A\mu+\epsilon^{AB}\mathscr D_B\upsilon$, then the angular momentum is 
                \begin{equation}
                    \label{eq-rt}
                    \mathcal J_\upsilon[\mathscr C]=-\frac{\varphi_0}{8\pi G_0}\oint_{\mathscr C}\upsilon\epsilon^{AB}\mathscr D_AN_B\ud^2\boldsymbol{\Omega},
                \end{equation}
and the Lorentz boost charge 
                \begin{eqnarray}
                    \label{eq-bt}
                    \mathcal K_\mu[\mathscr C]=-\frac{\varphi_0}{8\pi G_0}\oint_{\mathscr C}&&\mu\Bigg( \mathscr D^AN_A+2um\nonumber\\
                    &&-\frac{\hat c_A^{B}\hat c^A_B}{16}-\frac{2\omega+3}{8}\frac{\varphi_1^2}{\varphi_0^2} \Bigg)\ud^2\boldsymbol{\Omega}.
                \end{eqnarray}
In the above relations, $\mu$ and $\upsilon$ are linear combinations of $Y_{lm}$ with $l=1$.
The corresponding flux is
            \begin{eqnarray}
                \label{eq-f-y}
                    F_Y[\mathscr B]&=&F_{\alpha'}[\mathscr B]+\frac{\varphi_0}{32\pi G_0}\int_{\mathscr B}   Y^A\Bigg[ \frac{1}{2}(\hat c_B^C\mathscr D_AN^B_C-N_B^C\mathscr D_A\hat c^B_C)\nonumber\\
                    &&+\mathscr D^B(N^C_B\hat c_{AC}-\hat c^C_B N_{AC})\nonumber\\ 
                    &&+\frac{2\omega+3}{\varphi_0^2}(N\mathscr D_A\varphi_1-\varphi_1\mathscr D_AN)\Bigg]\ud u\ud^2\boldsymbol{\Omega},
            \end{eqnarray}
where $\alpha'=u\psi/2$. 
The flux-balance law in this case is 
\begin{equation}
    \label{eq-fb-y}
F_Y[\mathscr B]=-(\mathcal Q_Y[\mathscr C']-\mathcal Q_Y[\mathscr C]),
\end{equation}
with $\mathcal Q_Y$ either $\mathscr J_\upsilon$ or $\mathscr K_\mu$.

The flux-balance laws are very useful to constrain various memories, which will be discussed in the next section.

\section{Memories}
\label{sec-mem}

Finally, the memory effect is discussed.
Let us first use the obtained metric for an isolated system to compute the geodesic deviation equation, since the ground-based interferometer uses it to detect GWs.
Let $T^a$ be the 4-velocity of the test particles, and $S^a$ be the deviation vector.
Then one has (\cite{Wald:1984rg})
\begin{equation}
    \label{eq-gde}
    T^c\nabla_c(T^b\nabla_bS^a)=-R_{cbd}{}^aT^cS^bT^d.
\end{equation}
Assume the test particles are placed at a fixed radius $r_0$ and the fixed directions $x^A_0$, so these particles are called BMS detectors (\cite{Strominger:2014pwa}).
In general, these particles will be accelerated, but as long as $r_0$ is very large, they approximately freely fall.
One then has $T^a\approx(\pd_u)^a$ at a far distance $r_0$.
Define orthogonal spatial vectors $(e_{\hat \theta})^a=r^{-1}(\pd/\pd \theta)^a$, and $(e_{\hat\phi})^a=(r\sin\theta)^{-1}(\pd/\pd\phi)^a$. 
Then, let $S^a=S^{\hat A}(e_{\hat A})^a$, and the above equation becomes
\begin{equation}
    \label{eq-gde-c}
    \ddot S^{\hat A}\approx-R_{u\hat Bu}{}^{\hat A}S^{\hat B},
\end{equation}
where the electric part of the Riemann tensor $R_{abc}{}^d$ is 
\begin{equation}
    \label{eq-el-rie}
    R_{u\hat Au\hat B}=-\frac{1}{2r}\left( \pd_u^2\hat c_{\hat A\hat B}-\delta_{\hat A\hat B}\frac{\pd_u^2\varphi_1}{\varphi_0} \right)+\order{\frac{1}{r^2}}.
\end{equation}
Now, integrating eq.~\eqref{eq-gde-c} twice results in (\cite{Hou:2020tnd,Tahura:2020vsa})
\begin{equation}
    \label{eq-dev-c}
    \Delta S_{\hat A}\approx\frac{1}{2r}\left( \Delta\hat c_{\hat A\hat B}-\delta_{\hat A\hat B}\frac{\Delta\varphi_1}{\varphi_0} \right)S_{0}^{\hat B}+\order{\frac{1}{r^2}},
\end{equation}
in which $S^{\hat B}_0$ is the initial deviation vector at the retarded time $u_0$ when there were no GWs, i.e., $N_{AB}(u_0,x^C)=0$ and $N(u_0,x^A)=0$.
A radiating isolated system will eventually settle down to a state in which no GWs can be emitted, and so $N_{AB}$ and $N$ vanish again. 
But  $S^{\hat A}$ may not return to its original value, that is, 
\begin{equation}
    \label{eq-def-mm}
    \Delta S_{\hat A}\ne0.
\end{equation}
If so, there exists a permanent change in the relative distances between test particles.
This is the displacement memory effect.
By Eq.~\eqref{eq-dev-c}, there are two contributions to the total displacement memory, one of which is from the tensor part $\Delta\hat c_{AB}$, and the other from the scalar part $\Delta\varphi_1$.
So we will discuss the two contributions separately in Sec.~\ref{sec-dis}.
Section~\ref{sec-spin} concentrates on the spin and CM memories, mainly the constraints.

For the following discussion, one writes
\begin{equation}
    \label{eq-c-dec}
    \hat c_{AB}=\left(\mathscr D_A\mathscr D_B-\frac{1}{2}\gamma_{AB}\mathscr D^2\right)\Phi+\epsilon_{C(A}\mathscr D_{B)}\mathscr D^C\Upsilon,
\end{equation}
where $\Phi$ is the electric part and $\Upsilon$ the magnetic part.

\subsection{Displacement memories}
\label{sec-dis}

The displacement memory in the tensor sector is very similar to the one in GR.
It is related to the vacuum transition caused by the supertranslation as in GR (\cite{Strominger:2014pwa}).
As in GR, a vacuum state in the tensor sector is the spacetime with  $\hat c_{AB}=(\mathscr D_A\mathscr D_B-\frac{\gamma_{AB}}{2}\mathscr D^2)\Phi$ for some function $\Phi(x^A)$, so $N_{AB}=0$ (\cite{Hou:2020tnd}).
This kind of state can be transformed by a supertranslation $\alpha$ to a state with $\hat c'_{AB}=(\mathscr D_A\mathscr D_B-\frac{\gamma_{AB}}{2}\mathscr D^2)\Phi'$ and $\Phi'=\Phi-2\alpha$.
Therefore, the new state is also a vacuum, but different from the original one.
This observation suggests that there are infinitely many vacua in the tensor sector, and the transition between any pair of them is induced by a supertranslation.
Usually, one chooses one of the vacua as the physical one, e.g., $\hat c_{AB}=0$.
An infinitesimal Lorentz transformation results in $\delta_Y\hat c_{AB}=\lie_Y\hat c_{AB}-\frac{\psi}{2}\hat c_{AB}$ which is not a vacuum state, but it preserves the physical vacuum, so the S-matrix and the soft theorem are still Lorentz covariant.
The above observation suggests a similar explanation for the displacement memory to the one in GR. 
A process in question starts with an initial state which is a vacuum described by $\hat c_{AB}$, then the tensor GW penetrates $\scri$ and eventually disappears. 
So the final state is again a vacuum, which is likely a new one $\hat c'_{AB}$. 
The difference 
\begin{equation}
    \label{eq-def-dism}
    \Delta\hat c_{AB}=\hat c'_{AB}-\hat c_{AB}=\left(\mathscr D_A\mathscr D_B-\frac{\gamma_{AB}}{2}\mathscr D^2\right)\Delta\Phi,
\end{equation}
or equivalently, $\Delta\Phi=\Phi'-\Phi$, measures the memory effect.
This is nothing but
\begin{equation}
    \label{eq-dism-st}
    \Delta\Phi=-2\alpha,
\end{equation}
i.e., the displacement memory is induced by a certain supertranslation.

The displacement memory in the scalar sector, or S memory, is new.
Let the vacuum in the scalar sector be described by $\varphi_1=\varphi_1(x^A)$ and $N=0$.
Then one finds out that  a supertranslation does not transform it, i.e., $\delta_\alpha\varphi_1=0$, but a Lorentz generator $Y^A$ changes it according to $\delta_Y\varphi_1=\lie_Y\varphi_1+\frac{\psi}{2}\varphi_1$.
It is interesting to find out that the new state $\varphi'_1=\varphi_1+\delta_Y\varphi_1$ is also a vacuum state ($\pd_u\varphi'_1=0$).
Therefore, like the displacement memory in the tensor sector, the displacement memory $\Delta\varphi_1$ in the scalar sector is also the vacuum transition due to the Lorentz transformation.

The flux-balance law Eq.~\eqref{eq-fb-a} can be used to constrain the displacement memory in the tensor sector. 
In terms of the variation in $\Phi$, one has 
\begin{equation}
    \label{eq-f-al-2}
    \oint_{\mathscr C}\alpha\mathscr D^2(\mathscr D^2+2)\Delta\Phi\ud^2\boldsymbol{\Omega}=\frac{32\pi G_0}{\varphi_0}(\mathscr E_\alpha+\Delta\mathcal P_\alpha).
\end{equation}
Here, $\mathscr E_\alpha$ is Eq.~\eqref{eq-falpha} without the first term in the square brackets, that is, it is the null energy fluxes of the tensor and the scalar GWs.
The displacement memory in the scalar sector can be constrained by the evolution equation Eq.~(\ref{eq-uA2-ndot}), i.e.,
\begin{eqnarray}
    \label{eq-con-sm}
    \Delta\varphi_1^2=\frac{16\varphi_0^2}{2\omega+3}&\bigg\{&\frac{1}{32}\Delta(\hat c_A^B\hat c_B^A)+ \mathscr D^{-2}\mathscr D^A\Delta N_A \nonumber\\
    &&-\int_{u_i}^{u_f}\ud u\left[ m+\frac{1}{2}\mathscr D^{-2}\mathscr D^AJ_A \right]\bigg\},
\end{eqnarray}
where
\begin{equation}
    \label{eq-def-j}
    J_A=\frac{1}{2} N^B_C\mathscr D_A\hat c^C_B-\frac{2\omega+3}{\varphi_0^2}N\mathscr D_A\varphi_1,
\end{equation}
and $\mathscr D^{-2}$ is the inverse operator of $\mathscr D^2$ and is explicitly given in \cite{Hou:2020tnd}.

From the above discussion, one understands that indeed, the displacement memory in the tensor sector is very similar to the one in GR. 
For example, the both effects are due to the vacuum transition caused by the supertranslation in the tensor sector, i.e., Eq.~\eqref{eq-dism-st} (\cite{Strominger:2014pwa,Hou:2020tnd}).
At the same time, Eq.~\eqref{eq-f-al-2} is also similar to the constraint in GR.
For instance, Eq.~(3.11a) in \cite{Compere:2019gft} is the equation for constraining the displacement memory in GR based on the flux-balance law. 
However, this equation is a surface integral equation. 
To obtain the similar form to Eq.~\eqref{eq-f-al-2}, one simply integrates it over $u$.
Note that in that expression, $\Delta$ corresponds to $\mathscr D^2$ here,  $T$ is $\alpha$, and $\mathscr P_T$ is $\mathscr P_\alpha$.
Finally, I do not consider matter fields, so one may set $\hat T_{uu}$ in Eq.~(3.11a) to zero, for the purpose of comparison.
There are also two main differences between the displacement memories in BD and that in GR. 
First, the displacement memory effect in the tensor sector has a new contribution, i.e., the null energy flux of the scalar GW contained in $\mathscr E_\alpha$ (the last term in Eq.~\eqref{eq-falpha}), which is absent in GR.
Second, the S memory Eq.~\eqref{eq-con-sm} never appears in GR, because there does not exist a scalar degree of freedom  in GR.

\subsection{Spin and center-of-mass memories}
\label{sec-spin}

As mentioned in the Introduction, the spin memory can be detected by observing two counter-orbiting particles in a circle.
If there are GWs, they will return to their starting points at different times, given that they were released at the same time.
This effect is related to the leading term in $g_{uA}$ component (\cite{Pasterski:2015tva}), which does not depend on the scalar field. 
So the spin memory exists only in the tensor sector, and is very similar to the one in GR.

As in GR, the spin memory should be constrained by the flux-balance law associated with $Y^A$.
However, for this purpose, one allows $Y^A$ be a local conformal Killing vector field for $\gamma_{AB}$.
Then $Y^A$ has a finite number of singular points on the unit 2-sphere, and the flux-balance law obtained in Sec.~\ref{sec-ccf} should be modified, for example, keeping the charges Eq.~(\ref{eq-rt}) and (\ref{eq-bt}) while adding to Eq.~(\ref{eq-f-y}) the following term,
\begin{eqnarray}
    \label{eq-f-c}
    \Delta F_{Y}[\mathscr B]&=&\frac{\varphi_0}{32\pi G_0}\int_{\mathscr B}Y^A\mathscr D^B(\mathscr D_A\mathscr D_C\hat c^C_B-\mathscr D_B\mathscr D_C\hat c^C_A)\ud u\ud^2\boldsymbol{\Omega}\nonumber\\
        &=&\frac{\varphi_0}{64\pi G_0}\int_{\mathscr B}\epsilon_{AB}Y^A\mathscr D^B\mathscr D^2(\mathscr D^2+2)\Upsilon\ud u\ud^2\boldsymbol{\Omega}.
\end{eqnarray}
Then, the constraint on the spin memory, measured by $\Delta\mathcal R=\int\ud u\Upsilon$ (\cite{Flanagan:2015pxa}), reads,
\begin{equation}
    \label{eq-f-sc}
    \oint_{\mathscr C}\upsilon\mathscr D^2\mathscr D^2(\mathscr D^2+2)\Delta\mathcal R\ud^2\boldsymbol{\Omega}=-\frac{32\pi G_0}{\varphi_0}(\Delta\mathcal J_\upsilon+\mathscr Q_\upsilon+\mathscr J_\upsilon),
\end{equation}
where one defines
\begin{subequations}
    \begin{gather}
        \mathscr Q_\upsilon=-\frac{\varphi_0}{16\pi G_0}\int_{\mathscr B}\upsilon\epsilon_{AB}N^{AC}\hat c^B_C\ud u\ud^2\boldsymbol{\Omega},\\
        \mathscr J_\upsilon=\frac{\varphi_0}{16\pi G_0}\int_{\mathscr B}\upsilon\epsilon^{AB}\mathscr D_AJ_B\ud u\ud^2\boldsymbol{\Omega}.\label{eq-def-ju}
    \end{gather}
\end{subequations}
Note that here, $\upsilon$ is generally not a linear combination of $l=1$ spherical harmonics.

Finally, consider the constraint on the CM memory.
One may split $\Phi=\Phi_n+\Phi_o$ such that 
\begin{subequations}
    \begin{gather}
    \oint_{\mathscr C}\alpha\mathscr D^2(\mathscr D^2+2)\Delta\Phi_n\ud^2\boldsymbol{\Omega}=\frac{32\pi G_0}{\varphi_0}\mathscr E_\alpha,\\
    \oint_{\mathscr C}\alpha\mathscr D^2(\mathscr D^2+2)\Delta\Phi_o\ud^2\boldsymbol{\Omega}=\frac{32\pi G_0}{\varphi_0}\Delta\mathcal P_\alpha.
    \end{gather}
\end{subequations}
Then the CM memory effect is measured by (\cite{Nichols:2018qac,Tahura:2020vsa})
\begin{equation}
    \label{eq-def-cmm}
    \Delta\mathscr K=\int_{u_i}^{u_f}u\pd_u\Phi_o\ud u,
\end{equation}
which is contained in $F_{\alpha'}[\mathscr B]$ in Eq.~\eqref{eq-f-y}, i.e.,
\begin{equation}
    \label{eq-pp-k}
    F_{\alpha'}[\mathscr B]=-\frac{\varphi_0}{64\pi G_0}\oint_{\mathscr C}\mu\mathscr D^2\mathscr D^2(\mathscr D^2+2)\Delta\mathscr K\ud^2\boldsymbol{\Omega}.
\end{equation}
So the flux-balance law Eq.~\eqref{eq-fb-y} can be used to obtain,
\begin{equation}
    \label{eq-f-y-mu}
    \oint_{\mathscr C}\mu\mathscr D^2\mathscr D^2(\mathscr D^2+2)\Delta\mathscr K\ud^2\boldsymbol{\Omega}=\frac{64\pi G_0}{\varphi_0}(\mathscr J_\mu-\Delta\mathcal K'_\mu),
\end{equation}
where one defines
\begin{equation}
    \label{eq-def-kp}
    \Delta\mathcal K'_\mu=-\frac{\varphi_0}{8\pi G_0}\oint_{\mathscr C}\mu\Delta(\mathscr D^AN_A+2um)\ud^2\boldsymbol{\Omega}.
\end{equation}
Therefore, the CM memory is constrained by Eq.~\eqref{eq-f-y-mu}, as long as $\mu$ is not simply a linear combination of $l=1$ spherical harmonics.
Since CM memory is related to $\Phi_o$, which is a part of $\hat c_{AB}$, then there does not exist an analogous effect in the scalar sector.

\section{Conclusion}
\label{sec-con}

This work shows that the memory effect of an isolated system in BD shares some similarities with that in GR, and at the same time, has its own unique features. 
Both theories share the displacement memory effect for the tensor GW, which is induced by the passage of the null energy fluxes through $\scri$, because the supertranslations transform the degenerate vacua among each other in the tensor sector.
However, in BD, there exists the scalar GW.
It not only contributes to the tensor displacement memory effect by providing a new energy flux, but also has its own memory effect. 
The S memory effect is due to the angular momentum fluxes penetrating $\scri$,
and degenerate vacua in the scalar sector are transformed to each other by Lorentz transformations.
Using flux-balance laws, one obtains the constraints on the displacement memories, the spin memory and the CM memory.

Since the displacement memories in the tensor and the scalar sectors cause the permanent changes in the configuration of the interferometer's arms, the basic ideas for detecting them should be similar to those for detecting GR's displacement memory effect (\cite{Lasky:2016knh,McNeill:2017uvq,Johnson:2018xly,Hubner2020mmn,Boersma:2020gxx}).
It is also possible to use pulsar timing arrays to detect them as in GR, since the memory waveform generated by a supermassive binary black hole system can be modeled as a step function of time (\cite{Seto:2009nv,Wang2015mm}).
As in GR, the spin memory effect in BD should also be detected by LISA (\cite{Audley:2017drz,Nichols:2017rqr}), and likewise, Tianqin (\cite{Luo:2015ght}) or Taiji (\cite{Taiji2017}).
But the CM memory would be difficult to be detected by the current and even the planed detectors (\cite{Nichols:2018qac}).
The smoking gun would be the S memory effect.
The constraints from the flux-balance laws are very useful to predict how strong  memory effects are, and estimate whether they can be detected by the interferometer or pulsar timing arrays. 
Memory effect might provide a new method to test the nature of gravity.


\section*{Acknowledgments}

This work was supported by the \fundingAgency{National Natural Science Foundation of China} under grant Nos.\fundingNumber{11633001 and 11920101003}, and the \fundingAgency{Strategic Priority Research Program of the Chinese Academy of Sciences}, grant No.~\fundingNumber{XDB23000000}.
This was also a project funded by \fundingAgency{China Postdoctoral Science Foundation} (No.~\fundingNumber{2020M672400}).








\end{document}